\documentclass[a4paper,12pt]{article}

\newcommand{\be}{\begin{equation}}
\newcommand{\ee}{\end{equation}}
\newcommand{\bea}{\begin{eqnarray}}
\newcommand{\eea}{\end{eqnarray}}
\newcommand{\ba}{\begin{array}}
\newcommand{\ea}{\end{array}}
\newcommand{\req}[1]{~(\ref{#1})}

\newcommand{\nl}{\newline}
\newcommand{\hepth}[1]{{\tt hep-th/#1}}

\def\p{\partial}

\begin{document}
\begin{titlepage}

\begin{flushright}
UUITP-09/01\\
hep-th/0111245
\end{flushright}

\vspace{1cm}

\begin{center}
{\huge\bf  D2-brane RR-charge on $SU(2)$}
\end{center}
\vspace{5mm}

\begin{center}

{\large 
Peter Rajan} \\

\vspace{5mm}

Institutionen f\"or Teoretisk Fysik, Box 803, SE-751 08
Uppsala, Sweden

\vspace{5mm}

{\tt
peter.rajan@teorfys.uu.se
}

\end{center}

\vspace{5mm}

\begin{center}
{\large \bf Abstract}
\end{center}
\noindent
We compute RR charges of D2-branes on a background
with $H$-field which belongs to a nontrivial cohomology
class. We discover that the RR charge depends on the
configuration of the background `electric' RR field.
This result explains the ambiguity in the definition
of the RR charge previously observed in
the SU(2) WZW model.
\vfill
\begin{flushleft}
\end{flushleft}
\end{titlepage}
\newpage
\section{Introduction}
Computation of RR charges of D2 branes in the SU(2)
WZW model attracted considerable attention since this is
the simplest example where we deal with D-branes on
a curved background. Moreover, the field strength $H = dB$
of the $B$-field in this model is given by the volume form
on the 3-sphere, and, hence, belongs to a nontrivial
cohomology class. The first calculation by Bachas, Douglas
and Schweigert \cite{bach} produced the set of charges with
irrational ratios contradicting the standard idea of charge
quantization. A resolution of this paradox was proposed by Taylor
\cite{tayl} by taking into account the contribution 
of bulk fields into the RR charge.  In the case when $H$
belongs to a nontrivial cohomology class the Taylor's
mechanism produces an integral but ambiguous answer
 \cite{AleScho2},\cite{stanc}. For instance, it is argued
that in the $ SU(2)$ WZW model at level $k$ which describes
string propagation on a 3-sphere, the charge is defined
modulo $k$ \cite{stanc} or modulo $(k+2)$ \cite{AleScho2}. 
This ambiguity finds a mathematical interpretation in
terms of twisted K-theory (see \cite{SchFr},\cite{Freed},\cite{Mald1},\cite{Mald2}).
In this paper we reconsider the Taylor's argument in the
case of nontrivial $H$-field and discover that the 
RR charge depends on the background configuration
of the electric components of the RR field strength.
On a compact manifold, such as a 3-sphere, the RR field
equations possess nontrivial solutions even in the absence
of external sources. Adding such a free solution to 
the background configuration  changes
the RR charge of the brane. Hence, neglecting this
extra contribution leads to the charge ambiguity.

We start by exploring the analogy between D-brane physics 
and Maxwell's electromagnetism outlined by Taylor.
We first analyse a simple model of an expanding dipole
on the circle which then provides an intuition for 
computing RR charges of D2 branes on a 3-sphere.
\section{Electromagnetic analogy}
Recall the dimensional reduction in electromagnetism.
We start with the Maxwell's theory in the space-time dimension
$D=m+1$.  The fields in our theory are the gauge field $A=A_\mu dx^\mu$ 
and the (background) metric $g_{\mu \nu}$. The electromagnetic
field interacts with the external current represented by an
$m$-form $J=-*\tilde{J}$, such that $\tilde{J}= - \rho dt + {\bf j} \cdot d {\bf x}$
with $\rho$ density of the electric charge and ${\bf j}$ density
of the electric current. The action is given by formula,
\begin{equation}
S= \int 
\left[\frac{1}{2}*F\wedge F + A\wedge J\right] ,
\end{equation}
where $F=dA$ is the field strength of $A$.

Now assume that the $m$th spacial direction is compactified 
on a circle of length $l_m$, and  that the field
configuration is  always independent of the
$m$:th coordinate $x_m$. We can then dimensionally reduce the system
to $D'=(m-1)+1$ dimensions. The $D=(m+1)$ metric decomposes into the
$D'=(m-1)+1$ metric, the 1-form $g_m\sim g_{\mu m}$ and the 0-form
$g_{mm}$. Similarly, the fields $A$, $F$ and $J$
decompose as follows, \nl \nl
The 0-form $A_m$ \nl 
The 1-form $dA_m=F_m$\nl 
The 1-form $A$\nl 
The 2-form field strength $F=dA$\nl 
The (m-2)-form dual field strength $*F$\nl 
The (m-1)-form current $J$ \nl
\nl
In terms of dimensionally reduced variables the action
reads,
\begin{equation}
S = l_m \, \int \left[ \frac{1}{2}*F\wedge F   +   A\wedge J +
g_m\wedge\left(F_m\wedge *F + A_m J\right) + 
g_{mm} \left(F_m\wedge *F_m\right)\right]
\end{equation}
Since the energy momentum tensor $T$ is defined as $T^{\mu
\nu}=\frac{\delta S}{\delta g_{\mu\nu}}$ we find that
the momentum in the $m$:th direction is given by formula,
\be
P_m=\int_{V} T_{m0} = l_m \, 
\int_V \left[F_m\wedge *F + A_m J\right]
\ee
with $V$ implying integration over all of space.
It is convenient to split $P_m$ into two parts,
$$
p_m= l_m \int_V A_m J \, , \,
p'_m= l_m \int_V F_m\wedge *F,
$$
where, roughly speaking, $p_m$ is the momentum carried by the
electric current $J$, and $p'_m$ is the momentum contained in the
field configuration.

In what follows, $\Omega$ will be the region of space-time
contained between two hyper-planes $x^0 = 0$ and $x^0=T$,
and we assume that the fields decay sufficiently fast at 
the spatial infinity so as we can integrate by parts.
Over the time $T$, the momenta $p_m$ and $p'_m$ evolve as follows,
\be\label{EMmom}
\delta p_m=l_m \, \int_\Omega d\left(A_m J\right)=
l_m \, \int_\Omega F_m\wedge J
\ee
\be
\delta p'_m=l_m \, \int_\Omega d\left(F_m\wedge *F\right).
\ee
Momentum conservation follows from the Maxwell's equation,
$d * F  = J$, 
\be 
F_m\wedge J + d\left(F_m\wedge *F\right) =
F_m \wedge (J -  d *F) =0
\ee 
implying $\delta P_m = \delta p_m + \delta p'_m =0$.

Following Taylor we consider the following special situation:
let $J$ be produced by a slowly expanding dipole consisting
of charges $\pm q$. One first separates the charges from each other,
and then very slowly moves the charge $+q$ along some  path C.
The corresponding current $J$ is given by a $\delta$-function
supported on $C$, $J = - q \delta_C$. Then,
$$
\delta p_m = l_m \, \int_\Omega dA_m \wedge J =
- q l_m \int_C dA_m = - q l_m (A_m^f - A_m^i),
$$
where $A_m^{i,f}$ are the values of $A_m$ at the end-points
of $C$. Of course, $\delta p'_m = - \delta p_m$ to ensure
momentum conservation.

The situation becomes somewhat more interesting when 
in addition to $x^m$ one more direction is compactified.
For instance, consider a $2+1$-dimensional configuration 
where both spacial directions are compactified. The dimensional
reduction on the $m$th (now $m=2$) direction yields a
$1+1$-dimensional system where the spacial coordinate
is compactified on a circle of length $l_1$.

Again, we consider an expanding dipole, but now after the charge $+q$
makes a full turn around the circle, it can annihilate again with
 the charge $-q$. As before, this gives
\be 
\delta p_m= -ql_m (A_m^f - A_m^i) .
\ee 
Note that 
$$
l_m (A_m^f - A_m^i) = l_m\int_{S_1} \, dx^1 (\partial_1 A_m) =
l_m\int_{S_1} F_m = \Phi \, ,
$$
where $S_1$ is the circle  spanned by the 1st direction and $\Phi$ is the flux of the field $F$
through the torus spanned by both spatial directions. We conclude that $\delta p_m = - q \Phi$.

Seemingly, we arrive at a paradox since in general 
$\delta p_m \neq 0$ while the initial and final charge
configurations are the same. The latter, however, does not
imply that the field configurations at $x^0=0$ and $x^0 =T$
are the same. Indeed,  while expanding the dipole we  create an
electric field between the charges. By Maxwell's equation,
$$
*F|^T_0= \int_0^T \, dx^0 \, \p_0 *F  =
\int_0^T \, J_0 = \int_0^T \, {\bf j} = \delta q,
$$
where ${\bf j}$ is the electric current and $\delta q$
is the charge that passed through the given point
of the circle during the period from $x^0=0$ to $x^0=T$.
This extra electric field changes the momentum stored
in the field configuration,
\be 
\delta p'_m = l_m \int_{S_1} F_m\wedge *F |_0^T= 
l_m \int_{S_1} F_m (*F(T) - *F(0)) = q l_m \int_{S_1} F_m =
q \Phi.
\ee 

We conclude that the field configuration (the electric field)
contains the information about the history of our system.
Moving a charge $n$ times around the circle produces a
uniform electric field proportional to $n$. This electric
field (together with the background magnetic field) carries
momentum, and, since the space is compact, the configuration
with uniform electric and magnetic fields does not decay.

\section{RR-charge of D2-branes on $SU(2)$}Our main interest in this paper is the RR-charge of D2-branes. Following the reasoning outlined in the previous section we will deduce the D0 RR-charge from the type IIA supergravity action and show that just as the momentum of the electromagnetic system couldn't be determined by the charge configuration alone D0-charge can only be unambiguously defined when taking into account non-trivial free field solutions that are stable even in the absence of sources.\nl\nl D2-branes are found in type IIA supergravity but their origin can be traced back to M-theory. From this perspective the D2-branes are membranes charged with respect to the 3-form $C$ and the RR charge is just one particular component of 11-dimensional momentum. The low-energy limit of
M-theory is 11-dimensional supergravity which can be dimensionally
reduced to type IIA supergravity by compactifying a spatial
direction $x^m$ on a circle. While doing so the 3-form C decomposes
into the RR 3-form $ C^{(3)}$ and the NS-NS 2-form B. Momentum in the
m-direction is interpreted as D0-brane charge since D0-branes carry
charge under the R-R 1-form $C^{(1)}_\mu\sim g_{\mu m}$. From these fields one obtains the 3-form H=dB and the 4-form $G^{(4)}=dC^{(3)}-C^{(1)}\wedge
H$. The fields interact with external sources (D2-branes) through
the 7-form $J_{D2}$ defined so that for any 3-form $A^{(3)}$%
\be\int_{\Omega}J_{D2}\wedge A^{(3)}=\mu_2\int_{\Omega(\Sigma)}A^{(3)}\ee
 where $\Omega(\Sigma)$ is the world-volume of the brane $\Sigma$ and $\mu_2$ is a unit D2-brane charge. 
With kinetic terms canonically normalized (setting the prefactor $\frac{1}{2\kappa^{10}}$ to unity) the IIA supergravity action is (where ... denotes terms not involving
the field $C^{(1)}$): 
\bea
S_{IIA}&=&\int_{\Omega}\left\{\frac{1}{2}G^{(4)}\wedge
*G^{(4)}+...\right\}\nonumber\\&=&\int_{\Omega}
\left\{-C^{(1)}\wedge H\wedge *G^{(4)}+...\right\} \eea
The D2-brane world-volume action is a sum of a Born-Infeld action and a WZW
action. Since the Born-Infeld action does not involve the field $C^{(1)}$we ignore it in our considerations and focus on the WZW part which is:
\be\label{sd2wzw}
S^{D2}_{WZW}=\mu_2\int_{\Omega(\Sigma)}\left\{C^{(3)}+C^{(1)}\wedge\mathcal{F}\right\}=\int_{\Omega}\left\{J_{D2}\wedge C^{(3)}+J_{D2}\wedge C^{(1)}\wedge\mathcal{F}\right\} \ee
with $\mathcal{F}=B+2\pi\alpha'F$. $F$ is the 2-form field strength of the $U(1)$ gauge field on the brane.
Equations of motion give
$d*G^{(4)}=J_{D2}+H\wedge G^{(4)}$ i.e D2-branes are sources of the 'electrical' $*G^{(4)}-B\wedge G^{(4)}$ field. (\cite{bach},\cite{Zhou}) With no D4-branes we have $dG^{(4)}=0$.\nl\nl
Collecting terms that couple to $C^{(1)}$ we find the total D0-charge to be
$$
Q_{D0}=-\int_V \left\{*G^{(4)}\wedge H + J_{D2}\wedge\mathcal{F}
\right\}.
$$
Again, it's convenient to split $Q_{D0}$ into two parts; one associated with $J_{D2}$ and the other with the field configuration:
\be\label{d0charge} Q^{D2}_{D0}=-\int_V J_{D2}\wedge\mathcal{F} , \;\;Q^{field}_{D0}=-\int_V
*G^{(4)}\wedge H. \ee
The D0-charge evolves in time as follows:
\be
\delta Q^{D2}_{D0}=-\int_\Omega d(J_{D2}\wedge \mathcal{F})
\ee  
\be\label{elfieldcharge}
\delta Q^{field}_{D0}=-\int_\Omega d(*G^{(4)}\wedge H)
\ee
Charge conservation properties follows from
\bea\label{d0cons}\delta Q_{D0}&=& -\int_\Omega \left\{d(J_{D2}\wedge
\mathcal{F})+d(*G^{(4)}\wedge H)\right\}\nonumber\\&=&-\int_\Omega \left\{
d\left(J_{D2}\wedge (B+2\pi\alpha'F)+*G^{(4)}\wedge
H\right)-H\wedge G^{(4)}\wedge
H\right\}\nonumber\\&=&-\int_\Omega \left\{d\left(J_{D2}\wedge B+H\wedge
G^{(4)}\wedge B+*G^{(4)}\wedge H\right)-2\pi\alpha' J_{D2}\wedge
dF\right\}\nonumber\\&=&-\int_\Omega \left\{dd(*G^{(4)}\wedge
B)-2\pi\alpha' J_{D2}\wedge
dF\right\}\nonumber\\&=&2\pi\alpha'\mu_2\int_{\Omega(\Sigma)}dF\eea
where we have used the property $H\wedge H=0$. Note that while $dF=0$ on any
specific brane, we wish to consider a process where we expand a
brane and in this context $dF$ should be understood as the change of
$F$ as we pass through different brane configurations.  Evidentily, in
this framework, charge is in general only conserved as long as we
don't expand or shrink the brane. However, from momentum conservation
in M-theory we know that the charge has to be conserved, so if we wish
to expand a brane we have to add charge from external sources.
Using \cite{dbprimer}
\be
\mu_p=(2\pi)^{-p}\alpha'^{-\frac{p+1}{2}}
\ee 
we find that the amount of charge we add when expanding a brane is
\be\label{deld0}
\delta Q_{D0}=\mu_0\int_{\Omega(\Sigma)}\frac{dF}{2\pi}.\ee
Now we turn to the special case of a $D2$-brane on the $SU(2)$ group
manifold. If we parametrize the manifold with $(\psi,\theta,\phi)$
where $\psi$ describes the ''latitude'' on the 3-sphere ($\psi=\pi$
and $\psi=0$ corresponding to $-e$ and $e$ respectively) and
$(\theta,\phi)$ parametrizes the 2-spheres corresponding to fixed
$\psi$, then the brane is a 2-sphere sitting at fixed $\psi=\frac{\pi
n}{k}$ with $n,k$ integers satisfying $0<n<k$ (\cite{AleScho},\cite{AleScho3}). In the
semi-classical limit $k\to\infty$ we have a continuum of allowed brane
configurations. \nl Imagine taking a $D2$-brane from $e$ and expanding
it all the way to $-e$. At the pole it becomes singular and decouples
from the $B$-field so that we can transport it back to $e$ without any
loss or gain in momentum. (Note that since we work within the semi-classical limit $k\to\infty$ we disregard quantum corrections that would change this picture.) This is the analog of taking the full circle
in the dipole case.  The net change in $Q_{D0}$ of one
such winding around $S^3$ is
\be 
\delta Q_{D0}=\mu_0\int_{S^3}\frac{dF}{2\pi}
\ee
In the WZW level k model on $SU(2)$ \cite{AleScho},\cite{gawed},\cite{felder} the $H$-field is given by the volume 3-form to be \cite{bach},
\be
H=2k\alpha'\sin^2\psi\sin\theta d\psi d\theta d\phi
\ee 
Solving $H=dB$ gives
\be
B=k\alpha'\left(\psi-\frac{\sin 2\psi}{2}\right)\sin\theta d\theta d\phi
\ee
which is a well defined choice except at the point $\psi=\pi$. Another choice, $B'(\psi)=-B(\pi-\psi)$ is well defined everywhere except at the point $\psi=0$. The two choices are related by a gauge transformation. Gauge invariance of $\mathcal{F}=B+2\pi\alpha'F$ yields
\be
F=-\frac{k\psi}{2\pi}\sin\theta d\theta d\phi\;\;\mbox{or}\;\;F'=-F(\pi-\psi).
\ee 
Evaluating the integral we find 
\be
\delta Q_{D0}=-\mu_0 k
\ee
This is just the $mod$ k ambiguity of RR-charge of $D2$-branes on
$SU(2)$. Now we consider a general scenario where we first expand a membrane from pole-to-pole (as we did above) $\lambda$ times and then make an extra expansion under which the membrane traces out a 3-volume $\Gamma$ and ends up as the membrane $\Sigma$. Then, from \req{deld0}: 
\be\label{deld0su2}
\delta Q_{D0}=\mu_0\left(\int_\Sigma \frac{F(\Gamma)}{2\pi}-\lambda k\right)
\ee
with $F(\Gamma)$ being $F$ as defined on $\Gamma$. Since we started out with no charge, $\delta Q_{D0}$ should equal our expression (\ref{d0charge}) for the total charge $Q_{D0}=Q_{D0}^{D2}+Q_{D0}^{field}$. Using the expression following the first equality in (\ref{sd2wzw}) (rather then the second as we did to arrive at (\ref{d0charge})) we find:
\be
Q_{D0}^{D2}=\mu_2\int_{\Sigma} \mathcal{F}
\ee
While expanding the brane we also create a $*G^{(4)}$ field and the charge stored in this created field together with the $H$ field gives a contribution 
\bea
Q^{field}_{D0}&=&-\int_{\Omega}d\left(*G^{(4)}\wedge H\right)=-\int_{\Omega}J_{D2}\wedge H\nonumber\\&=&-\lambda\mu_2\int_{S^3}H-\mu_2\int_\Gamma H=-\mu_2\int_\Gamma H-\lambda\mu_0 k
\eea
Let $B(\Gamma)$ and $F(\Gamma)$ be choices of $B$ and $F$ that are  well defined everywhere on $\Gamma$, then
\be\label{qd0}
Q_{D0}=\int_{\Sigma}\mu_2\left(B+2\pi\alpha'F\right)-\mu_2\int_\Sigma B(\Gamma)-\lambda\mu_0 k=\mu_0\left(\int_\Sigma \frac{F(\Gamma)}{2\pi}-\lambda k\right).
\ee
As expected this is identical to \req{deld0su2}. In order for this formula to make sense we need to relate $\lambda$ and $\Gamma$ to physical properties of our configuration rather than referring to the history of it:
Let $M$ be the (compact) spatial
directions transverse to $S^3$ and assume that the above described process of expanding a membrane takes place between $x^0=0$ and $x^0=T$. Then, at $x^0=T$ the flux of $*G^{(4)}-B\wedge G^{(4)}$ on $M$ at any given point $x$ on $S^3$ is
\bea\label{g4flux} &&\int_M \left(*G^{(4)}(x)-B(x)\wedge G^{(4)}(x)\right)= \int_{M\times[0,T]} dx^0
\p_0\left(*G^{(4)}(x)-B(x)\wedge G^{(4)}(x)\right)
\nonumber\\&=&\int_{M\times[0,T]}d\left(*G^{(4)}(x)-B(x)\wedge G^{(4)}(x)\right)=\int_{M\times[0,T]}J_{D2}(x)=\int_\Omega J_{D2}\wedge\delta_{S^3}^{(3)}(x)\nonumber\\&=&\lambda\mu_2\int_{S^3}\delta_{S^3}^{(3)}(x)+\mu_2\int_{\Gamma}\delta_{S^3}^{(3)}(x)=\left(\lambda+\delta_\Gamma(x)\right)\mu_2 \eea
where we have used the equations of motion and compactness of $M$. In the above $\delta_{S^3}^{(3)}(x)$ is a deltafunction type 3-form on $S^3$ with support limited to the point $x$ and $\delta_\Gamma(x)$ satisfies
\be
\delta_\Gamma(x)=\left\{\begin{array}{ll} 1 & \textrm{if}\;x\in\Gamma\\0 & \textrm{if}\;x\notin\Gamma\end{array}\right..
\ee
The general expression \req{qd0} for $D0$ charge on $SU(2)$ is our main result. Since $\Gamma$ and $\lambda$ are determined by $\int_M\left(*G^{(4)}-B(x)\wedge G^{(4)}(x)\right)$ we find that $Q_{D0}$ is unambiguous. Also, integrality of $\int_{\Sigma}\frac{F}{2\pi}$ and $\lambda$ gives $Q_{D0}$ integral in units of $\mu_0$.
\section{Conclusions}
We have shown that ambiguities in the RR-charge on D-branes in B-field
backgrounds can be removed by taking into account nontrivial
configurations of the RR 'electric' field. This holds whenever all transverse directions $M$ are compact. If there are non-compact directions the electric field configuration is unstable since $\int_M \left(*G^{(4)}-B\wedge G^{(4)}\right)$ is no longer preserved. This allows charge to be transported to infinity and probably explains the charge ambiguity in the CFT approach.  
\section{Acknowledgments}
I would like to thank Anton Alekseev for his great support during the preparation of this paper.

\end{document}